\documentclass[10pt,conference]{IEEEtran}
\IEEEoverridecommandlockouts
\usepackage{booktabs}
\usepackage{multirow}
\usepackage{tabularx}
\usepackage{textcomp}
\usepackage{xcolor}
\usepackage{cite}
\usepackage{latexsym}
\usepackage{graphicx}
\usepackage{subcaption}
\usepackage{amsfonts,amssymb,amsmath}
\usepackage{nccmath}
\usepackage{optidef}
\usepackage{hyperref}
\usepackage{color,soul}
\usepackage{array}
\usepackage{algorithm}
\usepackage{algorithmic}
\usepackage[T1]{fontenc}
\usepackage[utf8]{inputenc}
\usepackage{fancyhdr}
\usepackage{lastpage}
\usepackage{caption}
\usepackage{soul}
\usepackage{subcaption}
\usepackage{setspace}
\usepackage{bm}
\usepackage{adjustbox}
\usepackage{makecell}
\usepackage{titlesec}
\def\BibTeX{{\rm B\kern-.05em{\sc i\kern-.025em b}\kern-.08em
    T\kern-.1667em\lower.7ex\hbox{E}\kern-.125emX}}
\usepackage{fancyhdr}
\fancypagestyle{firststyle}{
	\fancyhf{}
	\fancyhead[L]{H. Poddar and A. Chowdary "Single and Multi-Frequency Path Loss Models for Indoor Hotspot Scenario Based on Measurements Conducted at 6.75, 16.95, 28, 73 and 142 GHz'', \textit{in 2025 IEEE Global Communications Conference (GLOBECOM)}, Taipei, Taiwan, December 2025, pp. 1--6.} 
}
\allowdisplaybreaks
\begin{document}
\title{
Single and Multi-Frequency Path Loss Models for Indoor Hotspot Scenario Based on Measurements Conducted at 6.75, 16.95, 28, 73 and 142 GHz
}
\author{\IEEEauthorblockN{Hitesh Poddar$^{*}$ and Akhileswar Chowdary$^{\dagger}$}
\IEEEauthorblockA{$^{*}$Sharp Laboratories of America, Vancouver, WA, USA, poddarh@sharplabs.com \\$^{\dagger}$NYU WIRELESS, New York University, Brooklyn, NY, USA, akhileswar.chowdary@nyu.edu}
}
\maketitle
\thispagestyle{firststyle}
\begin{abstract}
This paper presents a comprehensive derivation of single and multi-frequency large-scale path loss model parameters for the close-in (CI) free space reference distance, CI free space reference distance with cross-polarization (CIX), floating-intercept (FI), CI free space reference distance with frequency-dependent path loss exponent (CIF), CI free space reference distance with frequency-dependent path loss exponent and cross-polarization (CIFX), alpha-beta-gamma (ABG), and alpha-beta-gamma with cross-polarization (ABGX) models for specific frequencies and across frequency ranges of 7–24 GHz, 0.5–100 GHz, and 0.5–150 GHz. The analysis is based on extensive real-world measurements conducted by NYU WIRELESS at 6.75 GHz, 16.95 GHz, 28 GHz, 73 GHz, and 142 GHz, using a 1 GHz wideband time-domain based sliding correlation channel sounder in the indoor hotspot (InH) scenario in both line-of-sight (LOS) and non-line-of-sight (NLOS) channel conditions. Specifically, the derived CI, FI, and ABG path loss model parameters for 7–24 GHz and 0.5–100 GHz frequency ranges in this article were submitted in Third Generation Partnership Project (3GPP) to validate Technical Report (TR) 38.901 InH path loss models, as part of the release (Rel) 19 study on ``Channel Model Validation of TR 38.901 for 7–24 GHz.'' Furthermore, the results in this paper provide critical insights into understanding large-scale path loss, comparing different path loss models, and extending the path loss models standardized by 3GPP and ITU for the InH scenario, which is essential for advancing next-generation wireless systems.
\end{abstract}
\begin{IEEEkeywords}
3GPP, ITU, TR 38.901, 6G, path loss models.
\end{IEEEkeywords}
\section{Introduction}
The exponential growth in mobile data traffic continues unabated, driven significantly by emerging indoor wireless applications such as 8K ultra-high-definition streaming, augmented reality, and virtual reality. These high-throughput use cases demand substantially larger bandwidths, predominantly available in the upper mid-band (FR3) \cite{kang:2023:cellular-wireless-networks, jiang:2021:the-road-towards}, millimeter-wave (mmWave), and sub-Terahertz (THz) frequency ranges, which are key enablers of next-generation wireless communication systems. While channel models for sub-6 GHz and mmWave bands have been extensively studied and standardized by 3GPP in Rel-14 and earlier, comprehensive channel characterization and modeling in emerging 6G candidate bands such as FR3 and sub-THz are still in early phases \cite{shakya:2024:comprehensive-fr1, shakya:2024:propagation, shakya:2024:wideband-penetration, chen:2024:an-experimental}. 
\par The FR3 band (7–24 GHz), often called the ``golden band,'' has gained considerable attention for global 6G spectrum allocations due to its optimal balance between coverage and capacity \cite{bjornson:2024:enabling-6g-performance, semaan:2023:6g-spectrum-enabling}. The ITU World Radiocommunication Conference (WRC) in 2023 identified key frequency bands for future wireless use, including 4.4–4.8 GHz, 7.125–7.250 GHz, 7.75–8.40 GHz, and 14.8–15.35 GHz within the FR3 band \cite{ghosh:2024:world-radiocommunications, ghosh:2024:the-national-spectrum, semaan:2023:6g-spectrum-enabling, davidson:2024:national-spectrum}, as well as the D band (110–170 GHz), G band (140–220 GHz), and H/J bands (220–330 GHz) within the sub-THz spectrum \cite{ghosh:2024:world-radiocommunications}. With 6G standardization efforts expected to commence in 2026, major standardization bodies such as 3GPP are actively studying the FR3 band for initial 6G deployments, while the THz spectrum is anticipated to be addressed in later phases of standardization.
\par The 3GPP TR 38.901 channel model \cite{3GPP:2020:study-on-channel-model} spans the frequency range of 0.5–100 GHz and relied on interpolation from sub-6 and mmWave measurements \cite{maccartney:2015:inh-mmWave} to estimate propagation behavior in the FR3 band. To address this gap, 3GPP approved a Rel-19 study on “Channel model validation of TR 38.901 for 7–24 GHz” \cite{3gpp:2023:sid-channel-modelling, poddar2025overview}. This paper contributes to that effort and future initiatives by 3GPP, ITU, and other standardization bodies \cite{ghosh:2024:world-radiocommunications, nga:2024:channel-measurements-and-modeling, ghosh:2024:the-national-spectrum} by providing single and multi-frequency omnidirectional path loss model parameters for the CI, CIX, FI, CIF, CIFX, ABG, and ABGX path loss models. The derived path loss model parameters for different path loss models are based on extensive real-world measurement campaigns conducted by NYU WIRELESS at 6.75 GHz, 16.95 GHz\footnote{6.75 GHz and 16.95 GHz were selected due to equipment constraints and Federal Communications Commission (FCC) authorization for conducting indoor and outdoor measurements in New York City, USA \cite{shakya:2024:comprehensive-fr1}. The selected frequencies 6.75 GHz and 16.95 GHz are representative of the lower and upper ends of the 7–24 GHz band, offering meaningful insight into path loss behavior across this wideband \cite{poddar2025validation3gpptr38901}.}, 28 GHz, 73 GHz, and 142 GHz using a 1 GHz wideband time-domain based sliding correlation channel sounder \cite{shakya:2024:comprehensive-fr1,rappaport:2024:point-data-for-site-specific,maccartney:2015:inh-mmWave,ju:2021:millimeter-wave-and-sub-terahertz} in LOS and NLOS. While additional data points across the 7–24 GHz, 0.5–100 GHz, and 0.5–150 GHz frequency ranges would further improve the accuracy of estimating path loss model parameters for different path loss models, constraints related to time, experimental licensing, hardware capabilities, and available InH environments limited this study. The findings in this work represent one of several contributions submitted or to be submitted to 3GPP, ITU \cite{gsma:2024:the-road}, or other standardization bodies. Final path loss model parameters by the standardization bodies will be determined by consolidating results from all such contributions. The main contributions of this work are summarized as follows:
\begin{itemize}
    \item The single frequency omnidirectional FI path loss model parameters are derived for 142 GHz in the InH scenario in both LOS and NLOS. This aids in extending the standardized InH path loss models in 3GPP and ITU above 100 GHz.
    \item The parameters of the multi-frequency omnidirectional path loss model (CI, FI, CIX, CIF, CIFX, ABG, and ABGX) are presented for frequency ranges 7-24 GHz and 0.5 to 100 GHz in the InH scenario for both LOS and NLOS. These contribute directly to the 3GPP Rel-19 study \cite{poddar2025overview}.
    \item The parameters of the multi-frequency omnidirectional path loss model (CI, FI, CIF, and ABG) are provided for the frequency range of 0.5–150 GHz for the InH scenario for both LOS and NLOS. This analysis offers foundational insights for future extension of the standardized InH path loss models in 3GPP, ITU, and other standardization bodies above 100 GHz.
\end{itemize}
The remainder of this paper is structured as follows. Section \ref{sec:overviewMeas} describes the measurement campaigns conducted by NYU WIRELESS. Sections \ref{sec:sfplm} and \ref{sec:mfplm} present the single- and multi-frequency path loss models, respectively, along with the parameters of the derived path loss model for different path loss models. Section \ref{sec:conclusion} concludes the article with a summary of key findings, limitations, and directions for future research.
\section{Overview of Measurement}\label{sec:overviewMeas}
Omnidirectional path loss measurements for the InH scenario in both LOS and NLOS were conducted by NYU WIRELESS at 6.75 GHz, 16.95 GHz, 28 GHz, 73 GHz, and 142 GHz using a 1 GHz wideband sliding correlation-based channel sounder \cite{shakya:2024:comprehensive-fr1,rappaport:2024:point-data-for-site-specific,maccartney:2015:inh-mmWave,ju:2021:millimeter-wave-and-sub-terahertz} using co-polarized and cross-polarized antennas at the transmitter (TX) and receiver (RX). Measurements at 6.75 and 16.95 GHz \cite{shakya:2024:comprehensive-fr1,rappaport:2024:point-data-for-site-specific} were carried out at the NYU WIRELESS Research Center located at 370 Jay Street, Brooklyn, NY, across 20 pairs of TX-RX locations (7 LOS, 13 NLOS) with a TX-RX separation ranging from 13 to 97 m. TX and RX antennas were at a height of 2.4 and 1.5 m, respectively. Furthermore, measurements at 28 and 73 GHz \cite{maccartney:2015:inh-mmWave} were performed at a previous NYU WIRELESS facility located at 2 Metro Tech Center, Brooklyn, NY, with 48 pairs of TX-RX (10 LOS, 38 NLOS) and separation distances of TX-RX ranging from 3.9 to 45.9 m. TX and RX antennas were at a height of 2.5 m and 1.5 m, respectively. Moreover, the 142 GHz \cite{ju:2021:millimeter-wave-and-sub-terahertz} measurements were carried out in the same InH facility as the 28 GHz and 73 GHz measurements using TX and RX antenna heights of 2.5 and 1.5 m, respectively. A subset of TX-RX pairs from the 28 GHz measurements was used at 142 GHz, yielding 21 TX-RX pairs (9 LOS, 12 NLOS) with separation distances ranging from 3.9 m to 39.2 m.
\section{Single Frequency path loss Models}\label{sec:sfplm}
Single-frequency path loss models characterize signal attenuation over distance at a given frequency. The most widely adopted single-frequency path loss models are the CI and FI path loss models. The CI path loss model anchors large-scale signal attenuation to a physically meaningful reference distance \(d_{0}=1\) m and models the path loss using a single parameter, the path loss exponent (PLE) \(n\) and has demonstrated superior accuracy and robustness to measurement variability in a wide range of data sets from various propagation environments \cite{sun2016investigation}. The CI path loss model is mathematically expressed as
\begin{align}
&\mathrm{PL}^{\mathrm{CI}}(f,d) = \mathrm{FSPL}(f,d_{0}) + 10n \log_{10}\bigl(d/d_{0}\bigr) + X_{\sigma}^{\mathrm{CI}}, \label{eq:ci_model}\\
&\mathrm{FSPL}(f,d)[\mathrm{dB}] = 32.4 + 20\log_{10}(f),\\
&\mathrm{PL}^{\mathrm{CI}}(f,d) = 32.4 + 20\log_{10}(f) + 10n\log_{10}\bigl(d/d_{0}\bigr) + X_{\sigma}^{\mathrm{CI}}, \label{eq:ci_sim_model}
\end{align}
where $d_{0}$ is the reference distance, \(\lambda\) is the wavelength in m, $f$ is the frequency in GHz, FSPL is the free space path loss, $d$ is the 3D distance between the TX and RX and \(X_{\sigma}^{\mathrm{CI}}\sim\mathcal{N}(0,\sigma^{2})\) models shadow fading in dB. Similarly, the details of the CIX path loss model are given in \cite{maccartney:2015:inh-mmWave}. 
\par In contrast, the FI path loss model does not incorporate a physically based reference point and instead models the path loss using two empirically derived parameters without any physical constraint, and is given by
\begin{equation}\label{eq:fi_model}
  \mathrm{PL}^{\mathrm{FI}}(d)
  = \alpha
    + 10\,\beta\,\log_{10}(d)
    + X_{\sigma}^{\mathrm{FI}},
\end{equation}
where \(\alpha\) (dB) is a floating-point intercept, and slope \(\beta\) is obtained by minimizing $\sigma$, and \(X_{\sigma}^{\mathrm{FI}}\sim\mathcal{N}(0,\sigma^{2})\) denotes shadow fading. Unlike the single-parameter CI path loss model, the FI path loss model requires two parameters, \(\alpha\) and \(\beta\), to optimally fit the measured data.
\par The 3GPP TR 38.901 path loss models for the InH scenario in LOS and NLOS are presented in \cite{3GPP:2020:study-on-channel-model, poddar2025validation3gpptr38901}. The 3GPP InH LOS path loss model [\eqref{eq:ci_sim_model},\cite{poddar2025validation3gpptr38901}] is mathematically equivalent to the CI path loss model in \eqref{eq:ci_sim_model}, with $\eta =$ 1.73. Furthermore, the 3GPP InH LOS path loss model is also equivalent to the FI path loss model in \eqref{eq:fi_model}, with parameters $\alpha = 32.4 + 20\log_{10}(f)$ (FSPL) and $\beta = 1.73$. \textit{This demonstrates that, for a single frequency, the 3GPP InH LOS path loss model is equivalent to both the CI and the FI path loss models}. Similarly, the 3GPP InH NLOS path loss model [Option 2, \eqref{eq:ci_sim_model}, \cite{poddar2025validation3gpptr38901}] corresponds to the CI path loss model in \eqref{eq:ci_sim_model} with $\eta =$ 3.19, and the FI path loss model with $\alpha = 32.4 + 20\log_{10}(f)$ and $\beta = 3.19$. However, the alternate option [Option 1, \eqref{eq:ci_sim_model}, \cite{poddar2025validation3gpptr38901}] is equivalent only to the FI path loss model with $\alpha = 17.3 + 24.9\log_{10}(f)$ and $\beta = 3.83$. \textit{This demonstrates that, for a single frequency, the 3GPP InH NLOS path loss model denoted by Option 2, \eqref{eq:ci_sim_model} in \cite{poddar2025validation3gpptr38901} is equivalent to the CI path loss model whereas Option 1, \eqref{eq:ci_sim_model} in \cite{poddar2025validation3gpptr38901} is equivalent to the FI path loss model.}

\begin{table}[t]
\caption{Single frequency omnidirectional CI and FI path loss model parameters at 6.75 GHz, 16.95 GHz, 28 GHz, 73 GHz, and 142 GHz for the InH scenario in LOS and NLOS with V-V antenna polarization.}
\resizebox{\columnwidth}{!}{
\begin{tabular}{|c|c|c|c|c|c|c|}
\hline
\multicolumn{1}{|c}{\multirow{2}{*}{\makecell{\textbf{Freq.}\\\textbf{(GHz)}}}} & \multicolumn{1}{|c}{\multirow{2}{*}{\textbf{Env.}}} & \multicolumn{2}{|c}{\textbf{CI: $\boldsymbol{d_0 = 1}$ m}} & \multicolumn{3}{|c|}{\textbf{FI}} \\
\cline{3-7}\multicolumn{1}{|c|}{} & & \textbf{PLE} & \makecell{$\boldsymbol{\sigma}$\\(\textbf{dB})} & \makecell{$\boldsymbol{\alpha}$\\(\textbf{dB)}} & \textbf{$\boldsymbol{\beta}$} & \makecell{$\boldsymbol{\sigma}$\\(\textbf{dB})} \\ 
\hline
\multirow{2}{*}{6.75} & LOS & 1.3 \cite{shakya:2024:comprehensive-fr1} & 3.5 & 43.4 & 1.7 \cite{poddar2025validation3gpptr38901} & 3.4 \\
\cline{2-7} & NLOS & 2.7 \cite{shakya:2024:comprehensive-fr1} & 9.2 & 35.2 & 3.6 \cite{poddar2025validation3gpptr38901} & 9.0 \\ 
\hline
\multirow{2}{*}{16.95} & LOS & 1.3 \cite{shakya:2024:comprehensive-fr1} & 2.7 & 50.9 & 1.7 \cite{poddar2025validation3gpptr38901}& 2.4 \\ 
\cline{2-7} & NLOS & 3.1 \cite{shakya:2024:comprehensive-fr1} & 8.1 & 61.0 & 2.8 \cite{poddar2025validation3gpptr38901} & 8.1 \\ 
\hline
\multirow{2}{*}{28} & LOS & 1.1 \cite{maccartney:2015:inh-mmWave} & 1.8 & 60.6 & 1.2 \cite{maccartney:2015:inh-mmWave} & 1.8 \\
\cline{2-7} & NLOS & 2.7 \cite{maccartney:2015:inh-mmWave} & 9.5 & 51.3 & 3.5 \cite{maccartney:2015:inh-mmWave} & 9.2 \\ 
\hline
\multirow{2}{*}{73} & LOS & 1.3 \cite{maccartney:2015:inh-mmWave} & 2.3 & 78.1 & 0.5 \cite{maccartney:2015:inh-mmWave} & 1.4 \\ 
\cline{2-7} & NLOS & 3.2 \cite{maccartney:2015:inh-mmWave} & 11.3 & 76.2 & 2.7 \cite{maccartney:2015:inh-mmWave}& 11.2 \\ 
\hline
\multirow{2}{*}{142} & LOS & 1.8 \cite{ju:2021:millimeter-wave-and-sub-terahertz} & 3.0 & \textbf{82.8} & \textbf{1.1} & \textbf{2.3} \\ 
\cline{2-7} & NLOS & 2.7 \cite{ju:2021:millimeter-wave-and-sub-terahertz} & 6.6 & \textbf{98.9} & \textbf{0.8} & \textbf{4.6} \\ 
\hline
\end{tabular}\label{tab:ci_fi_params}} 
\end{table}
\begin{figure}[!t]
    \centering
    \includegraphics[width=1.0\linewidth]{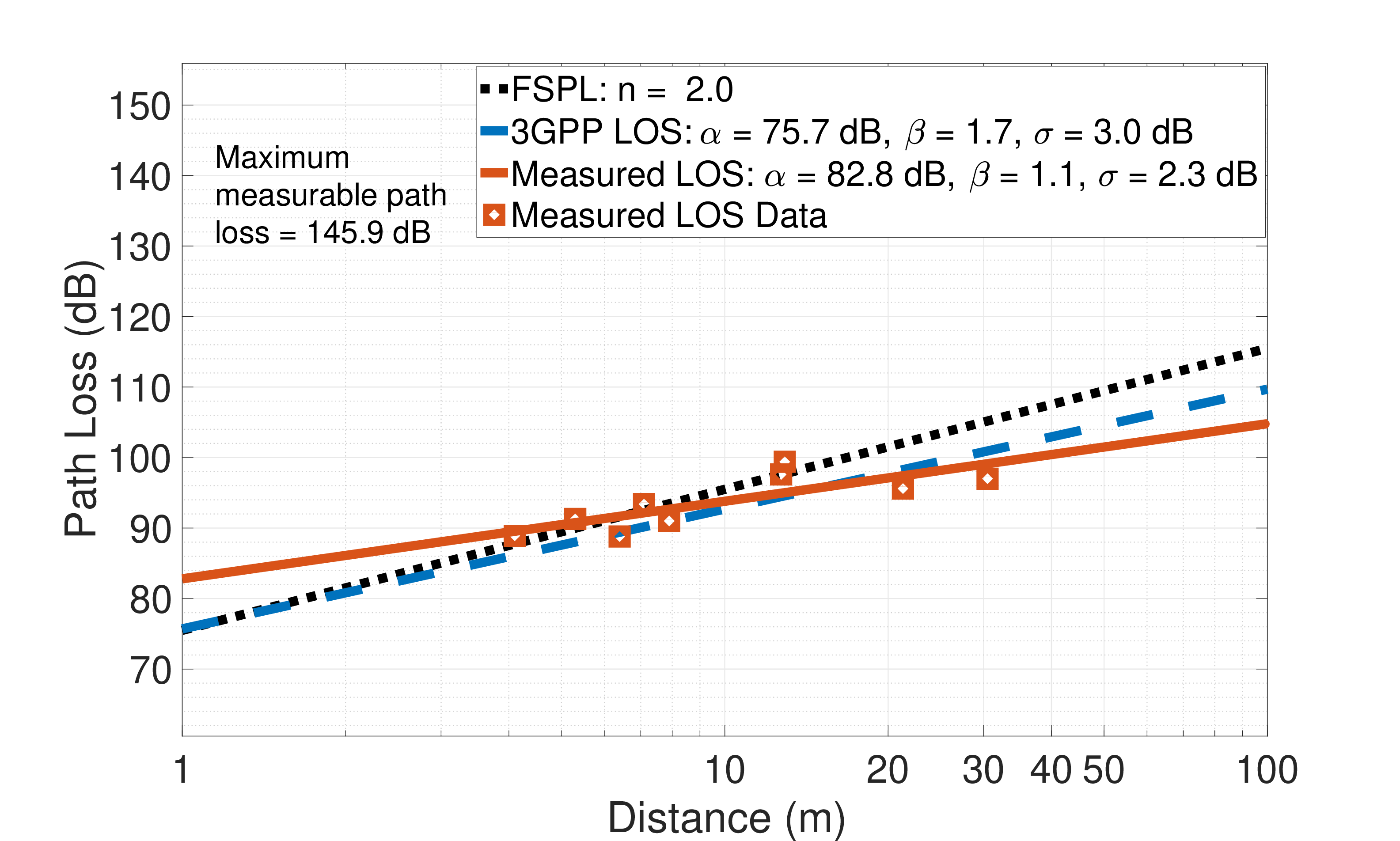}
    \caption{InH FI LOS path loss scatter plots and models for measured data and 3GPP path loss model over a distance range of 1-100 m for V-V polarization.}
    \label{fig:inhfimeas3gpplos}
\end{figure}
\par
\begin{figure}[!t]
    \centering
    \includegraphics[width=1.0\linewidth]{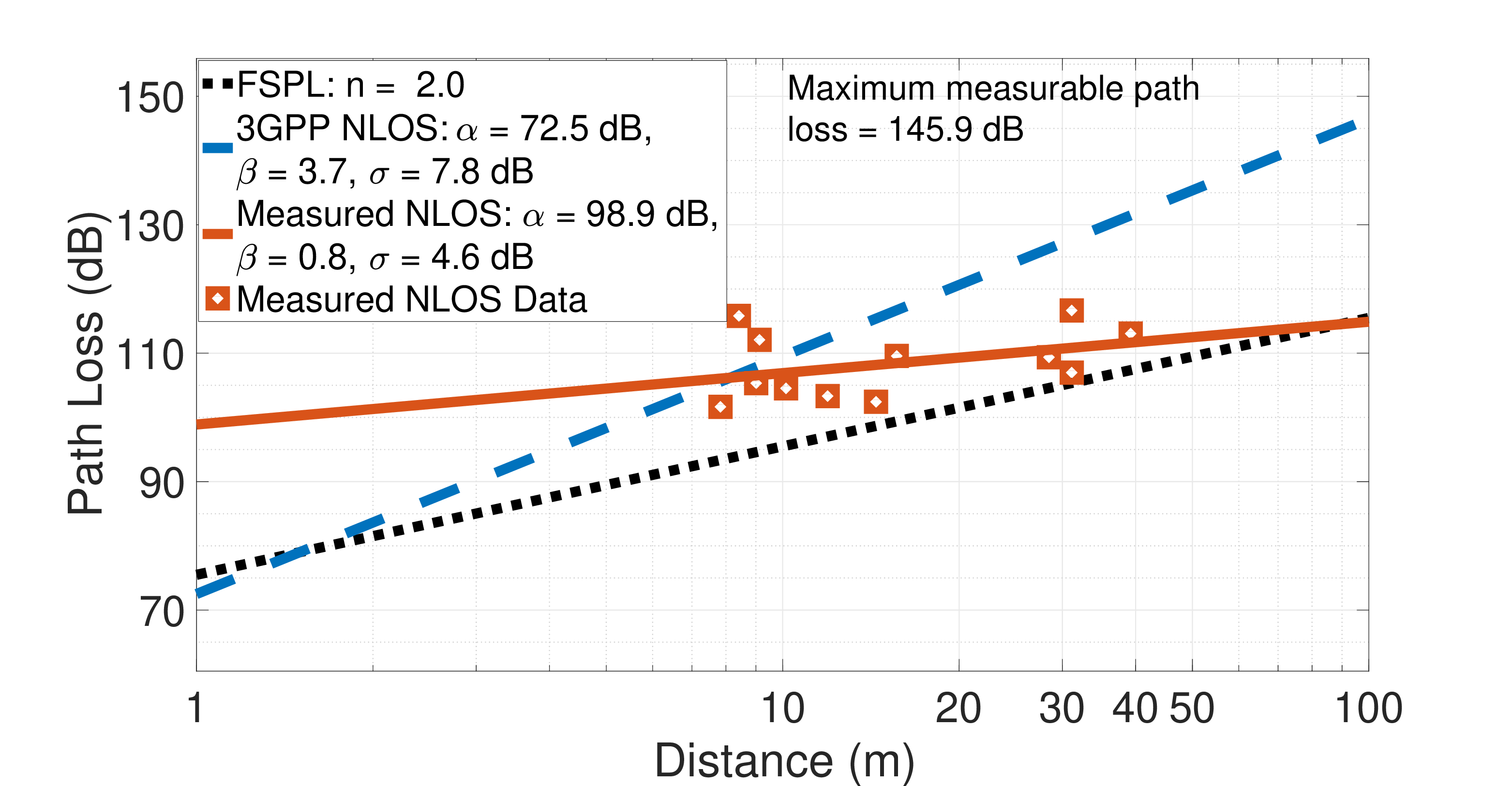}
    \caption{InH FI NLOS path loss scatter plots and models for measured data and 3GPP path loss model over a distance range of 1-100 m for V-V polarization.}
    \label{fig:inhfimeas3gppnlos}
\end{figure}

 Fig. \ref{fig:inhfimeas3gpplos} and Fig. \ref{fig:inhfimeas3gppnlos} present the fit of the FI path loss model for the measured data and the 3GPP TR 38.901 InH path loss model in both LOS and NLOS at 142 GHz. Table~\ref{tab:ci_fi_params} compares the parameters of the CI and FI path loss model between frequencies and provides the parameters of the FI path loss model at frequencies beyond 100 GHz. Table~\ref{tab:ci_fi_params} shows that the CI and FI path loss models yield nearly identical standard deviations for shadow fading at each frequency in the InH scenario in both LOS and NLOS. Furthermore, Table \ref{tab:ci_fi_params} reveals that the FI model exhibits significant variability in $\alpha$ compared to FSPL at 1 m across frequencies. In both LOS and NLOS, the $\alpha$ value fluctuates by nearly 40 dB and 60 dB, respectively, while the FSPL only varies by 25 dB (FSPL at 1 m at 6.75 GHz and 142 GHz is 49 dB and 75.5 dB, respectively) between 6.75 GHz and 142 GHz. This wide swing in $\alpha$ suggests that the FI model does not reliably reflect physical propagation conditions, especially in InH environments used during measurements, which had unobstructed free space for the first few meters. Furthermore, the $\beta$ values in the FI model show irregular and non-intuitive variations between frequencies. For example, although most frequencies in LOS yield relatively consistent $\beta$ values, there is a sharp drop at 73 GHz. A similar pattern is observed in NLOS, where the value $\beta$ for 142 GHz is the lowest among all frequencies. This anomalous result implies that higher frequencies experience less path loss than lower frequencies, which contradicts well-established propagation principles and defies physical intuition. The low $\beta$ values in the FI path loss model highlight its lack of physical anchoring. In the FI model, the intercept $\alpha$ captures not only the frequency-dependent FSPL at 1 m but also most of the frequency-dependent loss over distance, as seen in Fig. \ref{fig:inhfimeas3gppnlos} (the difference between the FSPL and $\alpha$ at 1 m is approximately 30 dB). Consequently, $\beta$ only captures any residual distance-dependent loss beyond what $\alpha$ has already captured, leading to non-physical values. Hence, in the FI model, the intercept $\alpha$ is not physically anchored to the FSPL at 1 m and may or may not align with it. Depending on the measurement data, $\alpha$ can be similar to the FSPL at 1 m or differ by tens of dB. This variability makes $\alpha$ and $\beta$ highly sensitive to data sets, and direct comparison across frequencies challenging.
In contrast, in the CI path loss model, the FSPL at 1 m is physically anchored. All additional distance-dependent losses beyond 1 m are captured by the PLE, which is equivalent to $\beta$ in the FI path loss model. Because the CI path loss model is anchored to a physical reference, which is the FSPL at 1 m, the PLE remains much more stable across a wide frequency range and less sensitive to the data sets compared to the FI path loss model. \textit{Thus, while 3GPP currently adopts the CI path loss model for LOS and treats it as optional for NLOS, future 3GPP standardization efforts could consider establishing the CI path loss model as the primary model for the InH scenario in both LOS and NLOS, given its physical grounding, stability, and consistency across frequencies compared to the FI path loss model.}

\section{Multi-Frequency path loss Models}\label{sec:mfplm}
The multi-frequency path loss models capture the dependence of signal attenuation on both distance and frequency. While the ABG path loss model is most commonly used for multi-frequency path loss modeling, the potential of the CI path loss model is often overlooked despite its physical grounding and parameter stability across a wide range of frequencies. The ABG model is mathematically expressed as
\begin{multline}\label{eq:ABG_model}
\text{PL}^{\text{ABG}}(f,d) [\text{dB}] = 10\alpha \log{10}\left(\frac{d}{d_0}\right) + \beta \\+ 10\gamma \log_{10}\left(\frac{f}{1 \text{GHz}}\right) + X^{\text{ABG}}_{\sigma},
\end{multline}
where $d_0 = 1 \text{ m}$ and $d$ is the 3D TX-RX distance. Here, $\alpha$ and $\gamma$ quantify the distance and frequency dependence of path loss, respectively. The parameter $\beta$ represents an optimized offset without direct physical interpretation, and $f$ denotes the frequency in GHz. $X^{\text{ABG}}_{\sigma} \sim \mathcal{N}(0,\sigma^2)$ denotes shadow fading, representing large-scale signal fluctuations around mean path loss as the distance varies.

\textit{The 3GPP InH path loss model for both LOS and NLOS [\eqref{eq:ci_sim_model} and \eqref{eq:fi_model} in \cite{poddar2025validation3gpptr38901}] are equivalent to the ABG path loss model in \eqref{eq:ABG_model} for multi-frequencies}. Specifically, the 3GPP LOS path loss model can be derived from the ABG path loss model in \eqref{eq:ABG_model} by setting the parameters $\beta = 32.4$, $\alpha = 1.73$, and $\gamma = 2$. Similarly, the 3GPP InH NLOS path loss model variants defined in \eqref{eq:fi_model} of \cite{poddar2025validation3gpptr38901} correspond to the ABG path loss model with parameter sets $\beta = 17.3, \alpha = 3.83, \gamma = 2.49$ for Option 1 in \eqref{eq:ci_sim_model}, \cite{poddar2025validation3gpptr38901}, and $\beta = 32.4, \alpha = 3.19, \gamma = 2$ for Option 2 in \eqref{eq:fi_model}, \cite{poddar2025validation3gpptr38901}. The ABG path loss model can revert to the CI or FI path loss model through specific parameter configurations. The ABG path loss model becomes equivalent to the CI path loss model when $\beta = 32.4$, $\alpha = n$ (PLE), and $\gamma = 2$. Likewise, the 3GPP ABG path loss model reverts to the FI path loss model when $\alpha = \beta + 10\gamma\log_{10}(f)$ and $\beta = \alpha$.

Moreover, a specific extension of the ABG path loss model, incorporating an XPD factor, is introduced for scenarios involving cross-polarized antennas at the TX and RX. The ABG path loss model with the XPD factor, also referred to as the ABGX path loss model, is described in \cite{maccartney:2015:inh-mmWave}.

\begin{table}[t]
\centering
\caption{Multi-frequency omnidirectional CI, CIX, CIF, CIFX, ABG, and ABGX path loss model parameters for 7-24 GHz based on measurements at 6.75 GHz and 16.95 GHz in the InH scenario in both LOS and NLOS.}
\resizebox{\columnwidth}{!}{
\begin{tabular}{|c|c|c|c|c|c|c|}
\hline
\multicolumn{7}{|c|}{\makecell{\textbf{LOS path loss Model Parameters}}} \\ 
\hline
\multicolumn{1}{|c}{} &\multicolumn{1}{|c}{\textbf{Pol.}} &\multicolumn{1}{|c}{\textbf{PLE}}&\multicolumn{2}{|c}{\textbf{XPD (dB)}}&\multicolumn{2}{|c|}{$\boldsymbol{\sigma}$ (\textbf{dB})} \\ 
\hline
CI & V-V &1.3 &\multicolumn{2}{c}{-} &\multicolumn{2}{|c|}{3.1}\\ 
\hline
CIX & V-H & 1.3 & \multicolumn{2}{c}{18.5} & \multicolumn{2}{|c|}{6.9}\\ 
\hline
\multicolumn{1}{|c}{} & \multicolumn{1}{|c}{} & \multicolumn{1}{|c}{$\boldsymbol{n}$}& \multicolumn{1}{|c}{$\boldsymbol{b}$}&\multicolumn{1}{|c}{$\boldsymbol{f_{0}}$ (\textbf{GHz})}&\multicolumn{1}{|c}{\textbf{XPD (dB)}}&\multicolumn{1}{|c|}{$\boldsymbol{\sigma}$ (\textbf{dB})}\\ 
\hline
CIF & V-V & 1.3 & 0 & 12.0 & - &3.1 \\ 
\hline
CIFX & V-H & 1.3 & 0 & 12.0 & 16.9 & 6.7 \\ 
\hline
\multicolumn{1}{|c}{} & \multicolumn{1}{|c}{} &\multicolumn{1}{|c}{$\boldsymbol{\alpha}$}& \multicolumn{1}{|c}{$\boldsymbol{\beta}$}&\multicolumn{1}{|c}{$\boldsymbol{\gamma}$}&\multicolumn{1}{|c|}{\textbf{XPD (dB)}} &\multicolumn{1}{|c|}{$\boldsymbol{\sigma}$ (\textbf{dB})} \\ 
\hline
ABG \cite{poddar2025validation3gpptr38901} & V-V & 1.7 & 28.2 & 1.9 & - & 2.9  \\ 
\hline
ABGX & V-H & 1.7 & 28.2 & 1.9 & 17.6  & 6.6 \\ 
\hline
\multicolumn{7}{|c|}{\makecell{\textbf{NLOS path loss Model Parameters}}} \\ 
\hline
\multicolumn{1}{|c}{} & \multicolumn{1}{|c}{\textbf{Pol.}}&\multicolumn{1}{|c}{\textbf{PLE}}&\multicolumn{2}{|c}{\textbf{XPD (dB)}}&\multicolumn{2}{|c|}{$\boldsymbol{\sigma}$ (\textbf{dB})} \\ 
\hline
CI & V-V & 2.9 &\multicolumn{2}{c}{-} &\multicolumn{2}{|c|}{9.1}\\ 
\hline
CIX & V-H & 2.9 & \multicolumn{2}{c}{15.8} & \multicolumn{2}{|c|}{11.9}\\ 
\hline
\multicolumn{1}{|c}{} &\multicolumn{1}{|c}{}& \multicolumn{1}{|c}{$\boldsymbol{n}$}& \multicolumn{1}{|c}{$\boldsymbol{b}$}&\multicolumn{1}{|c}{$\boldsymbol{f_{0}}$ \textbf{(GHz)}}&\multicolumn{1}{|c|}{\textbf{XPD (dB)}}&\multicolumn{1}{|c|}{$\boldsymbol{\sigma}$ (\textbf{dB})} \\ 
\hline
CIF & V-V & 2.9 & 0.1 & 12.0 & - & 8.7  \\ 
\hline
CIFX & V-H & 2.9 & 0.1 & 12.0 & 21.8 & 12.0  \\ 
\hline
\multicolumn{1}{|c}{} &\multicolumn{1}{|c}{}&\multicolumn{1}{|c}{$\boldsymbol{\alpha}$}& \multicolumn{1}{|c}{$\boldsymbol{\beta}$}&\multicolumn{1}{|c}{$\boldsymbol{\gamma}$}& \multicolumn{1}{|c|}{\textbf{XPD (dB)}} &\multicolumn{1}{|c|}{$\boldsymbol{\sigma}$ (\textbf{dB})} \\ 
\hline
ABG \cite{poddar2025validation3gpptr38901} & V-V & 3.2 & 12.9 & 3.4 & - & 8.6 \\ 
\hline
ABGX & V-H & 3.2 & 12.9 & 3.4 & 16.2  & 10.6  \\ 
\hline
\end{tabular}\label{tab:ci_abg_params_fr3}} 
\end{table}

\begin{table}[t]
\centering
\caption{Multi-frequency omnidirectional CI, CIX, CIF, CIFX, ABG, and ABGX path loss model parameters for 0.5-100 GHz based on measurements conducted at 6.75 GHz, 16.95 GHz, 28 GHz, and 73 GHz for the InH scenario in both LOS and NLOS with V-V antenna polarization.}
\resizebox{\columnwidth}{!}{
\begin{tabular}{|c|c|c|c|c|c|c|}
\hline
\multicolumn{7}{|c|}{\makecell{\textbf{LOS path loss Model Parameters}}} \\ 
\hline
\multicolumn{1}{|c}{} &\multicolumn{1}{|c}{\textbf{Pol.}}&\multicolumn{1}{|c}{\textbf{PLE}}&\multicolumn{2}{|c}{\textbf{XPD (dB)}}&\multicolumn{2}{|c|}{$\boldsymbol{\sigma}$ (\textbf{dB})} \\ 
\hline
CI & V-V & 1.3 &\multicolumn{2}{c}{-} &\multicolumn{2}{|c|}{2.8}\\ 
\hline
CIX & V-H & 1.3 & \multicolumn{2}{c}{18.0} & \multicolumn{2}{|c|}{6.2}\\ 
\hline
\multicolumn{1}{|c}{} & \multicolumn{1}{|c}{} & \multicolumn{1}{|c}{$\boldsymbol{n}$}& \multicolumn{1}{|c}{$\boldsymbol{b}$}&\multicolumn{1}{|c}{$\boldsymbol{f_{0}}$ (\textbf{GHz})}&\multicolumn{1}{|c}{\textbf{XPD (dB)}}&\multicolumn{1}{|c|}{$\boldsymbol{\sigma}$ (\textbf{dB})}\\ 
\hline
CIF & V-V & 1.3 & 0 & 35.0 & - & 2.8 \\ 
\hline
CIFX & V-H & 1.3 & 0 & 35.0 & 18.0 & 6.2 \\ 
\hline
\multicolumn{1}{|c}{} & \multicolumn{1}{|c}{} & \multicolumn{1}{|c}{$\boldsymbol{\alpha}$}& \multicolumn{1}{|c}{$\boldsymbol{\beta}$}&\multicolumn{1}{|c}{$\boldsymbol{\gamma}$}&\multicolumn{1}{|c|}{\textbf{XPD (dB)}} &\multicolumn{1}{|c|}{$\boldsymbol{\sigma}$ (\textbf{dB})} \\ 
\hline
ABG \cite{poddar2025validation3gpptr38901} & V-V & 1.4 & 29.5 & 2.1 & - & 2.7  \\ 
\hline
ABGX & V-H & 1.4 & 29.5 & 2.1 & 18.4 & 6 \\ 
\hline
\multicolumn{7}{|c|}{\makecell{\textbf{NLOS path loss Model Parameters}}} \\ 
\hline
\multicolumn{1}{|c}{} &\multicolumn{1}{|c}{\textbf{Pol.}}&\multicolumn{1}{|c}{\textbf{PLE}}&\multicolumn{2}{|c}{\textbf{XPD (dB)}}&\multicolumn{2}{|c|}{$\boldsymbol{\sigma}$ (\textbf{dB})} \\ 
\hline
CI & V-V & 2.9 &\multicolumn{2}{c}{-} &\multicolumn{2}{|c|}{10.5}\\ 
\hline
CIX & V-H & 2.9 & \multicolumn{2}{c}{13.8} & \multicolumn{2}{|c|}{10.8}\\ 
\hline
\multicolumn{1}{|c}{} & \multicolumn{1}{|c}{} &\multicolumn{1}{|c}{$\boldsymbol{n}$}& \multicolumn{1}{|c}{$\boldsymbol{b}$}&\multicolumn{1}{|c}{$\boldsymbol{f_{0}}$ (\textbf{GHz})}&\multicolumn{1}{|c|}{\textbf{XPD (dB)}}&\multicolumn{1}{|c|}{$\boldsymbol{\sigma}$ (\textbf{dB})} \\ 
\hline
CIF & V-V & 3.0 & 0.1 & 40.0 & - & 10.2  \\ 
\hline
CIFX & V-H & 3.0 & 0.1 & 40.0 & 16.2 & 10.9  \\ 
\hline
\multicolumn{1}{|c}{} & \multicolumn{1}{|c}{} &\multicolumn{1}{|c}{$\boldsymbol{\alpha}$}& \multicolumn{1}{|c}{$\boldsymbol{\beta}$}&\multicolumn{1}{|c}{$\boldsymbol{\gamma}$}& \multicolumn{1}{|c|}{\textbf{XPD (dB)}} &\multicolumn{1}{|c|}{$\boldsymbol{\sigma}$ (\textbf{dB})} \\ 
\hline
ABG \cite{poddar2025validation3gpptr38901} & V-V & 3.4 & 12.9 & 2.9 & - & 10.1 \\ 
\hline
ABGX & V-H & 3.4 & 12.9 & 2.9 & 13.9 & 10.3  \\ 
\hline
\end{tabular}\label{tab:ci_abg_params_fr3_fr2}} 
\end{table}

\begin{table}[t]
\centering
\caption{Multi-frequency omnidirectional CI, CIF, and ABG path loss model parameters for 0.5-150 GHz based on measurements conducted at 6.75 GHz, 16.95 GHz, 28 GHz, 73 GHz, and 142 GHz for the InH scenario in both LOS and NLOS with V-V antenna polarization. CIX, CIFX, and ABGX path loss model parameters could not be derived due to the lack of publicly available measurement data for the 142 GHz V-H polarization.}
\resizebox{\columnwidth}{!}{
\begin{tabular}{|c|c|c|c|c|c|}
\hline
\multicolumn{6}{|c|}{\makecell{\textbf{LOS path loss Model Parameters}}} \\ 
\hline
\multicolumn{1}{|c}{} & \multicolumn{1}{|c}{\textbf{Pol.}}& \multicolumn{3}{|c}{\textbf{PLE}}&\multicolumn{1}{|c|}{$\boldsymbol{\sigma}$ (\textbf{dB})} \\ 
\hline
CI & V-V & \multicolumn{3}{c|}{1.4} & 3.5\\ 
\hline
\multicolumn{1}{|c}{} & \multicolumn{1}{|c}{\textbf{Pol.}}& \multicolumn{1}{|c}{$\boldsymbol{n}$}& \multicolumn{1}{|c}{$\boldsymbol{b}$}&\multicolumn{1}{|c}{$\boldsymbol{f_{0}}$ (\textbf{GHz})}&\multicolumn{1}{|c|}{$\boldsymbol{\sigma}$ (\textbf{dB})}\\ 
\hline
CIF & V-V & 1.4 & 0.1 & 57.0 & 3.0 \\ 
\hline
\multicolumn{1}{|c}{} & \multicolumn{1}{|c}{\textbf{Pol.}}& \multicolumn{1}{|c}{$\boldsymbol{\alpha}$}& \multicolumn{1}{|c}{$\boldsymbol{\beta}$}&\multicolumn{1}{|c}{$\boldsymbol{\gamma}$}&\multicolumn{1}{|c|}{$\boldsymbol{\sigma}$ (\textbf{dB})} \\ 
\hline
ABG & V-V & 1.5 & 24.3 & 2.4 & 3.1  \\ 
\hline
\multicolumn{6}{|c|}{\makecell{\textbf{NLOS path loss Model Parameters}}} \\ 
\hline
\multicolumn{1}{|c}{} & \multicolumn{1}{|c}{\textbf{Pol.}}& \multicolumn{3}{|c}{\textbf{PLE}}&\multicolumn{1}{|c|}{$\boldsymbol{\sigma}$ (\textbf{dB})} \\ 
\hline
CI & V-V & \multicolumn{3}{c|}{2.9} & 10.7\\ 
\hline
\multicolumn{1}{|c}{} & \multicolumn{1}{|c}{\textbf{Pol.}}& \multicolumn{1}{|c}{$\boldsymbol{n}$}& \multicolumn{1}{|c}{$\boldsymbol{b}$}&\multicolumn{1}{|c}{$\boldsymbol{f_{0}}$ (\textbf{GHz})}&\multicolumn{1}{|c|}{$\boldsymbol{\sigma}$ (\textbf{dB})} \\ 
\hline
CIF & V-V & 2.9 & 0 & 51.0 & 10.2  \\ 
\hline
\multicolumn{1}{|c}{} & \multicolumn{1}{|c}{\textbf{Pol.}}& \multicolumn{1}{|c}{$\boldsymbol{\alpha}$}& \multicolumn{1}{|c}{$\boldsymbol{\beta}$}&\multicolumn{1}{|c}{$\boldsymbol{\gamma}$}&\multicolumn{1}{|c|}{$\boldsymbol{\sigma}$ (\textbf{dB})} \\ 
\hline
ABG & V-V & 3.1 & 23 & 2.5 & 10 \\ 
\hline
\end{tabular}\label{tab:ci_abg_params_all}} 
\end{table}

Additionally, a two-parameter multi-frequency model termed the CIF model, which maintains the physically meaningful anchor of the FSPL at 1 m, is described by
\begin{multline}
\text{PL}^{\text{CIF}}(f,d) [\text{dB}] = \text{FSPL}(f,d_0)\\ + 10n\left[1 + b\left(\frac{f - f_0}{f_0}\right)\right]\log{10}\left(\frac{d}{d_0}\right) + X^{\text{CIF}}_{\sigma},
\end{multline}
where $d_0 = 1 \text{ m}$, $d$ is the 3D TX-RX distance, $n$ denotes the baseline PLE, $b$ captures the linear frequency dependence of the path loss, and $f_0$ is a weighted reference frequency computed from the frequency distribution of the measurements. $X^{\text{CIF}}_{\sigma} \sim \mathcal{N}(0,\sigma^2)$ models shadow fading. When $b = 0$, the CIF model reverts to the simpler CI model, thereby emphasizing its versatility and practical applicability in various multi-frequency measurement scenarios. Similarly, the CIFX model provides an extension of the CIF model to account for cross-polarized measurements and is described in \cite{maccartney:2015:inh-mmWave}.

The parameters of the multifrequency omnidirectional path loss model for the 7-24 GHz, 0.5-100 GHz, and 0.5-150 GHz frequency ranges are presented in Tables~\ref{tab:ci_abg_params_fr3}, \ref{tab:ci_abg_params_fr3_fr2}, and \ref{tab:ci_abg_params_all}, respectively. Due to differences in measurement environments, system setups, and post-processing techniques, individual organizations may observe discrepancies between measured path loss values and those predicted by the standardized 3GPP or ITU path loss models. This variation is expected, as the 3GPP/ITU InH path loss models span a wide frequency range from 0.5-100 GHz, making it inherently difficult to fine-tune the path loss models for any one frequency or band without affecting accuracy across others. As a result, tuning the path loss model specifically for bands such as FR3, mmWave, or THz based solely on localized measurements risks compromising the path loss model consistency and predictive reliability across the broader frequency range. To maintain backward compatibility and preserve the integrity of existing standardized path loss models, updates should not be made in isolation based on individual data sets. Instead, data from multiple independent measurement campaigns, both historical and recent, should be aggregated to ensure that any refinements reflect a comprehensive and representative view of propagation behavior. This holistic approach is essential to ensure consistency in the path loss model throughout the 0.5-100 GHz frequency range \cite{poddar2025validation3gpptr38901}. As future standardization efforts begin to extend the path loss models for frequencies beyond 100 GHz, it is critical that the data used to develop these extensions include both sub-100 GHz and above-100 GHz path loss measurements. This will support the creation of unified and backward-compatible path loss models that are applicable over a wide frequency range, as illustrated in Table \ref{tab:ci_abg_params_all}.

From Tables~\ref{tab:ci_abg_params_fr3}, \ref{tab:ci_abg_params_fr3_fr2}, and \ref{tab:ci_abg_params_all}, we can observe that for LOS, minimal differences are observed among PLEs for the CI, CIX, CIF, CIFX, ABG, and ABGX path loss models. This is also true for NLOS, but the NLOS PLEs are higher compared to LOS PLEs because of higher path loss. The standard deviations for the CI, CIF, and ABG path loss models are similar, indicating that all three path loss models similarly capture the LOS and NLOS path loss characteristics effectively for the 7-24 GHz, 0.5-100 GHz, and 0.5-150 GHz frequency ranges, respectively. In particular, the CIF model in Tables~\ref{tab:ci_abg_params_fr3}, \ref{tab:ci_abg_params_fr3_fr2}, and \ref{tab:ci_abg_params_all}shows that the frequency weighting ($b$) is negligible for both LOS and NLOS, reflecting no frequency dependence. Similarly, in Tables~\ref{tab:ci_abg_params_fr3} and \ref{tab:ci_abg_params_fr3_fr2} for LOS, the CIX, CIFX, and ABGX path loss models present similar standard deviations but significantly higher standard deviations compared to the CI, CIF, and ABG path loss models in LOS, emphasizing substantial variation caused by polarization mismatches. This is also true for NLOS. However, for NLOS, the standard deviations for all path loss models notably increase compared to LOS because of the higher propagation complexity.

\begin{figure}[!t]
    \centering
    \includegraphics[width=1.0\linewidth]{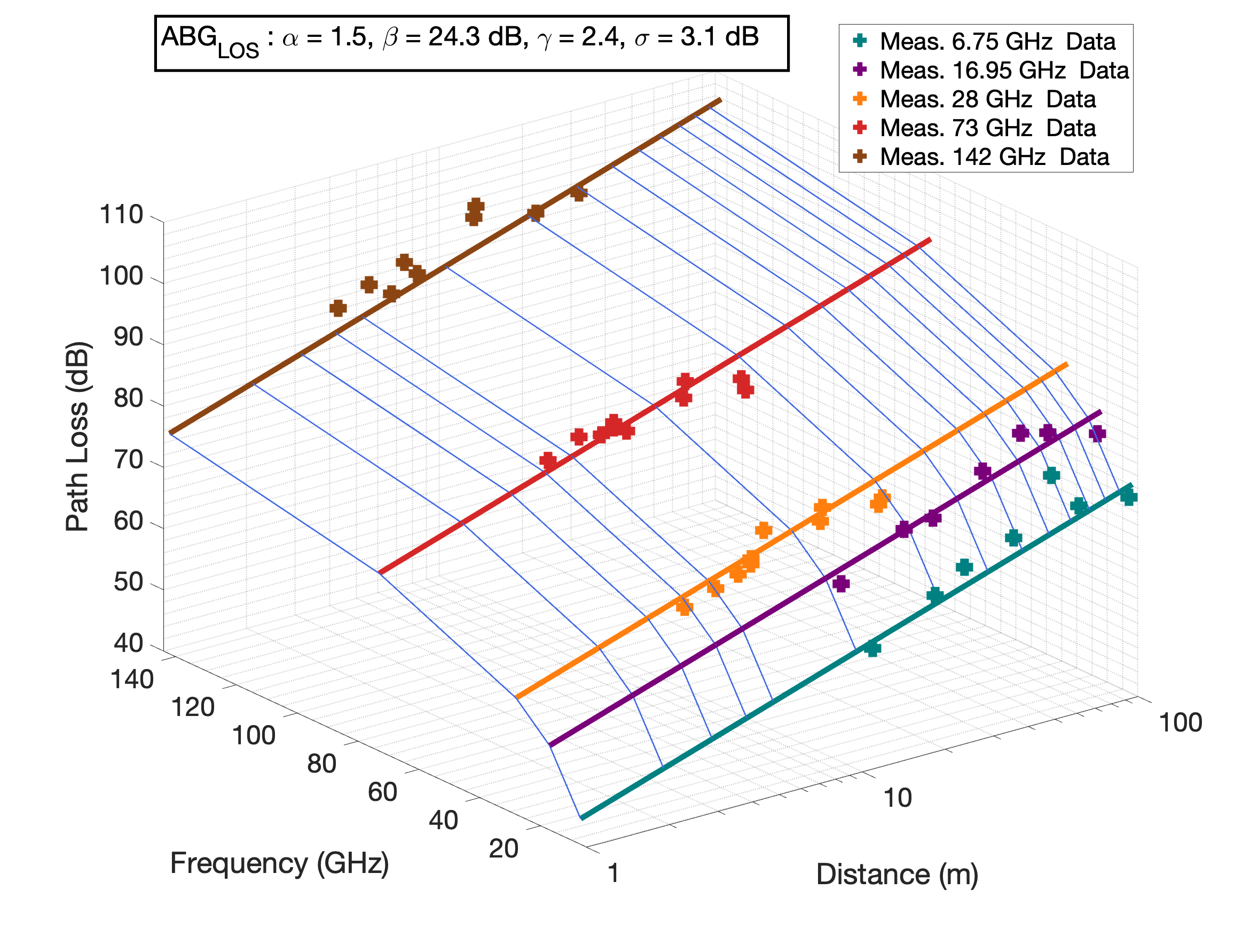}
    \caption{InH LOS ABG path loss scatter plots and models for 0.5-150 GHz over a distance range of 1 m to 100 m for V-V polarization.}
    \label{fig:inhabglos}
\end{figure}
\begin{figure}[!t]
    \centering
    \includegraphics[width=1.0\linewidth]{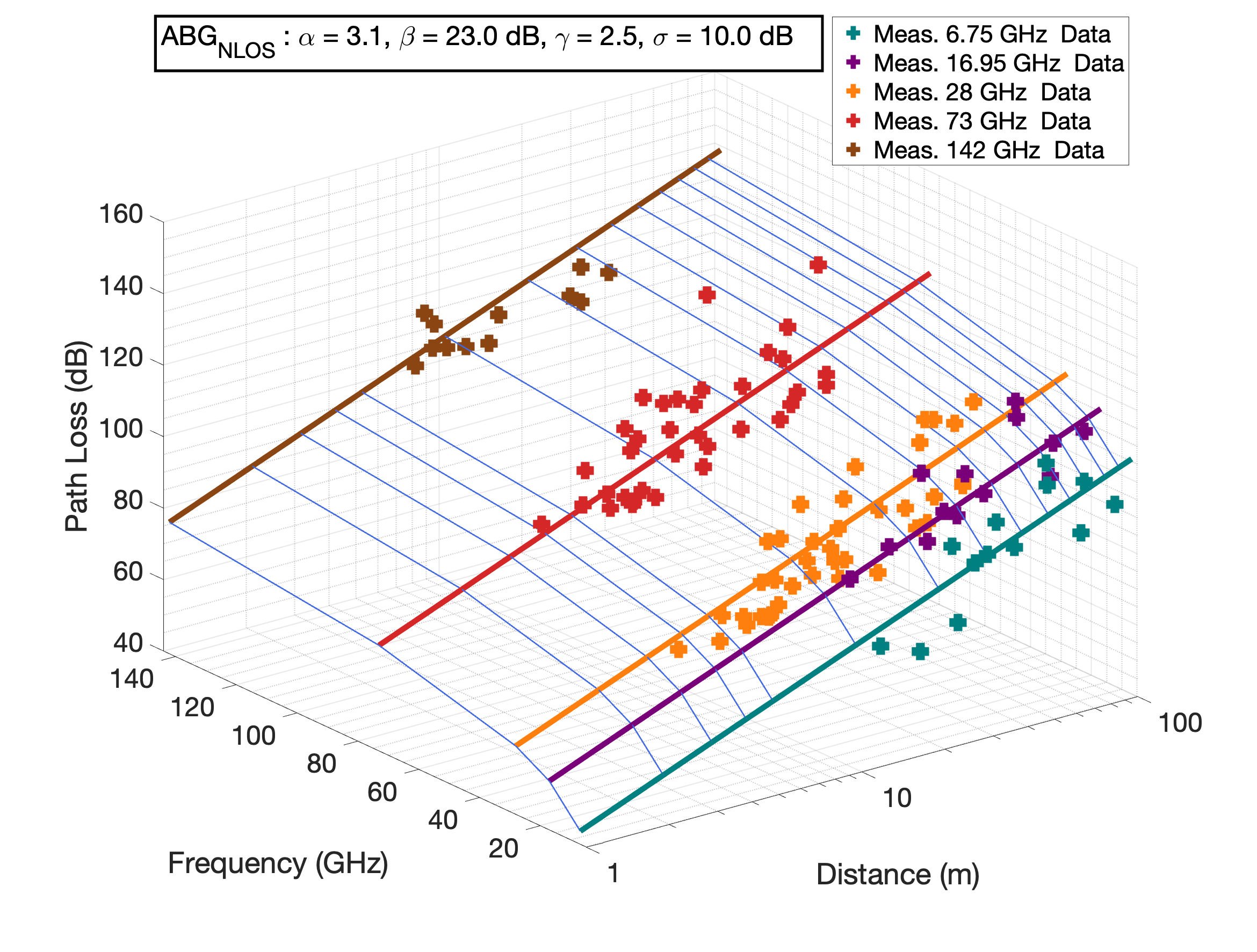}
    \caption{InH NLOS ABG path loss scatter plots and models for 0.5-150 GHz over a distance range of 1 m to 100 m for V-V polarization.}
    \label{fig:inhabgnlos}
\end{figure}

Figs. \ref{fig:inhabglos} and \ref{fig:inhabgnlos} present the ABG path loss model fit for the measured data in both LOS and NLOS for the InH scenario for the frequency range of 0.5-150 GHz. Tables~\ref{tab:ci_abg_params_fr3}, \ref{tab:ci_abg_params_fr3_fr2}, and \ref{tab:ci_abg_params_all} show that the CI path loss model exhibits stable PLE values of approximately 1.3 in LOS and 2.9 in NLOS in the 7-24 GHz, 0.5-100 GHz, and 0.5-150 GHz ranges. In contrast, the parameters of the ABG path loss model $\alpha$ and $\gamma$ show greater variability in different frequency ranges. Specifically, $\alpha$ remains relatively stable in different frequency ranges compared to $\gamma$ in both LOS and NLOS. This is mainly because $\alpha$ captures the path loss dependence on distance, and each measurement campaign is carried out at a specific frequency over a wide range of distances. Hence, $\alpha$ has multiple sampling points spread throughout the distance range of 1 to 100 m from various measurement campaigns. In contrast, $\gamma$ is limited by the number of frequency points in a wide frequency range (only two frequency points within the 7–24 GHz range, four in the 0.5-100 GHz range, and five in the 0.5-150 GHz spectrum). This sparse frequency sampling in these wide frequency ranges introduces significant variability in $\gamma$, making it sensitive to the specific measured data. Hence, incorporating additional frequency points from a variety of InH environments would improve the stability and accuracy of the parameter $\gamma$. \textit{These results support the recommendation that standardization bodies such as 3GPP and ITU could adopt the use of the CI path loss model for robust multi-frequency path loss modeling and simulations}. \par Furthermore, as stated in Section III, both the 3GPP InH LOS and NLOS (Option 2) path loss models are equivalent to the CI path loss model. As shown in Table \ref{tab:ci_abg_params_all}, the PLE obtained from the CI model is 1.4 for LOS and 2.9 for NLOS, which differs from the corresponding 3GPP-specified values of 1.73 and 3.19, respectively \cite{poddar2025validation3gpptr38901}. In both cases, the measured PLEs fall outside the 95\% confidence intervals (1.3–1.5 for LOS and 2.8–3.1 for NLOS), indicating statistically significant deviations. These discrepancies may be attributed to the waveguide effect commonly observed in indoor environments, where measurements were conducted across different frequencies in long, narrow hallways that enhance signal strength, resulting in stronger received power and reduced PLE. \textit{To further validate and refine the 3GPP InH LOS and NLOS CI path loss models, additional measurements are needed in different InH environments.} On the other hand, as stated in Section III, the 3GPP InH NLOS path loss model (Option 1) is equivalent to the ABG path loss model. As shown in Table \ref{tab:ci_abg_params_all}, the values of $\alpha$ and $\gamma$ obtained from the ABG model are 3.1 and 2.5, respectively, which differ from the corresponding 3GPP-specified values of 3.83 and 2.49 \cite{poddar2025validation3gpptr38901}. The measured $\alpha$ falls outside the 95\% confidence interval of 2.4 to 3.7, indicating a statistically significant deviation, while $\gamma$ is similar to the specified 3GPP value. The lower $\alpha$ may again be attributed to environment-specific wave guide effects. \textit{Hence, further measurements across varied InH environments are necessary to validate and refine the parameter $\alpha$ in the 3GPP ABG path loss model.}
\section{Conclusion}\label{sec:conclusion}
This study presents single- and multi-frequency omnidirectional path loss model parameters for the CI, FI, CIX, CIF, CIFX, ABG, and ABGX models in the InH scenario in both LOS and NLOS. The path loss model parameters for different path loss models are based on extensive real-world measurements conducted by NYU WIRELESS at 6.75 GHz, 16.95 GHz, 28 GHz, 73 GHz, and 142 GHz, covering key frequency ranges of interest for 6G and beyond, including FR3, mmWave, and sub-THz bands. Furthermore, this work demonstrates that the 3GPP TR 38.901 InH path loss models are mathematically equivalent to the CI, FI, and ABG models under appropriate parameter selections. The path loss parameters derived for the 7-24 GHz and 0.5-100 GHz frequency ranges have been submitted to 3GPP as part of the Rel-19 study on the validation of TR 38.901. Furthermore, the results for the 0.5-150 GHz frequency range offer valuable insights for future standardization efforts aimed at developing or extending the path loss models beyond 100 GHz. However, the limited number of frequency points, two within 7-24 GHz, four within 0.5-100 GHz, and five within 0.5-150 GHz, renders the derived path loss model parameters sensitive to the measured data and environment. As such, the robustness and generalization of the derived path loss models may be constrained. To address this, future work should incorporate additional frequency measurements in a broader set of indoor environments with varying structural and material characteristics. This will enable more accurate and comprehensive modeling, thereby supporting the extension and refinement of 3GPP and ITU path loss models for next-generation wireless systems.

\bibliographystyle{IEEEtran}

\end{document}